\documentclass[10pt,a4paper,twocolumn]{article}

\usepackage{lineno}

\usepackage{authblk}
\usepackage{graphicx}
\usepackage[explicit]{titlesec}
\usepackage[labelfont=bf,labelsep=endash,font=footnotesize]{caption}
\usepackage{subcaption}
\usepackage{xcolor}
\usepackage[
    backend=biber, 
    natbib=true,
    style=numeric,
    maxnames=50,
    sorting=none
]{biblatex}
\usepackage[hidelinks]{hyperref}
\usepackage[hang]{footmisc}
\usepackage[normalem]{ulem}
\usepackage[top=2.5cm, bottom=2.8cm, left=1.5cm, right=1.5cm]{geometry}
\usepackage{abstract}
\usepackage{mathtools}
\usepackage{balance}
\usepackage{wrapfig}
\usepackage{threeparttable}
\usepackage{makecell}
\usepackage{multirow}

\makeatletter
\renewcommand\AB@affilsepx{, \protect\Affilfont}
\makeatother

\providecommand{\keywords}[1]{\textbf{Keywords}\ \ \textendash\ \   #1}

\titleformat{\section}{\large\bfseries}{\thesection.}{1em}{\MakeUppercase{#1}}
\titlespacing*{\section}{0pt}{12pt}{6pt}

\titleformat{\subsection}{\large}{\thesubsection}{1em}{#1}
\titlespacing*{\subsection}{0pt}{12pt}{6pt}

\titleformat{\subsubsection}{\large\itshape}{\thesubsubsection}{1em}{#1}
\titlespacing*{\subsubsection}{0pt}{12pt}{6pt}

\newcommand{\ITUurl}[1]{\textcolor{black}{\urlstyle{same}\url{#1}}}

\newcommand{\ITUpar}{\vspace{8pt}\par}

\renewenvironment{abstract}
               {\list{}{
               \setlength{\rightmargin}{0mm}
               \setlength{\leftmargin}{0mm}
               \vspace{-0.25in}
                \item[\textit{\textbf{\hspace{22pt}Abstract  }}  \textendash]\relax}}
               {\endlist}

\def\starttable{\vspace{6pt}\begin{table}[ht]\center}
\def\startfigure{\vspace{6pt}\begin{figure}[ht]\center}

\makeatletter
\def\tagform@#1{\maketag@@@{\ignorespaces#1\unskip\@@italiccorr}}
\makeatother

\title{\large{\textbf{\uppercase{Multimodal Transformers for Wireless Communications: A Case Study in Beam Prediction}}}}

\author{\normalsize{Yu Tian}}
\author{\normalsize{Qiyang Zhao}}
\author{\normalsize{Zine el abidine Kherroubi}}
\author{\normalsize{Fouzi Boukhalfa}}
\author{\normalsize{Kebin Wu}}
\author{\normalsize{Faouzi Bader}}

\affil{\normalsize{Technology Innovation Institute, 9639 Masdar City, Abu Dhabi, United Arab Emirates}}

\date{\vspace{-12pt}{\small Corresponding author: Qiyang Zhao, qiyang.zhao@tii.ae.} \\  \endgraf\rule{\textwidth}{1pt}}

\begin{document}

\twocolumn[

\begin{@twocolumnfalse}
\maketitle

\begin{abstract}
\textit{Wireless communications at high-frequency bands with large antenna arrays face challenges in beam management, which can potentially be improved by multimodality sensing information from cameras, LiDAR, radar, and GPS. In this paper, we present a multimodal transformer deep learning framework for sensing-assisted beam prediction. We employ a convolutional neural network to extract the features from a sequence of images, point clouds, and radar raw data sampled over time. At each convolutional layer, we use transformer encoders to learn the hidden relations between feature tokens from different modalities and time instances over abstraction space and produce encoded vectors for the next-level feature extraction. We train the model on a combination of different modalities with supervised learning. We try to enhance the model over imbalanced data by utilizing focal loss and exponential moving average. We also evaluate data processing and augmentation techniques such as image enhancement, segmentation, background filtering, multimodal data flipping, radar signal transformation, and GPS angle calibration. Experimental results show that our solution trained on image and GPS data produces the best distance-based accuracy of predicted beams at 78.44\%, with effective generalization to unseen day scenarios near 73\% and night scenarios over 84\%. This outperforms using other modalities and arbitrary data processing techniques, which demonstrates the effectiveness of transformers with feature fusion in performing radio beam prediction from images and GPS. Furthermore, our solution could be pretrained from large sequences of multimodality wireless data, on fine-tuning for multiple downstream radio network tasks.}
\end{abstract}

\ITUpar

\ITUpar
\ITUpar

\end{@twocolumnfalse}
]

\section{Introduction} 
\label{sec:introduction}
Wireless communications beyond 5G is exploiting high-frequency bands such as mmWave and THz, in order to boost the system capacity by utilizing large available bandwidth. Massive antenna arrays have been leveraged to create ultra-narrow beams, so as to increase the received signal power and reduce interference on targeted users. Significant challenges in beam management arise in such systems and scenarios especially for high mobility vehicle users, to provide ultra-high reliable and low latency communications. 

Multimodality sensory information has the potential to improve wireless communications in a challenging environment. Integrated sensing and communication has been actively studied for 6G \cite{computer_vision}. In the vehicular network scenario, a roadside Base Station (BS) unit equipped with a camera, LiDAR, radar, and GPS can produce images, point clouds, radar signals, and location information of the road environment, objects, and vehicle users (UE). Such sensory data is potentially useful in assisting the BS to analyze the radio transmission scenario, so as to produce effective beam management. 
\subsection{\textcolor{black}{Problem statement}}
\begin{figure*}[ht]
\centering
\includegraphics[width= 16cm]{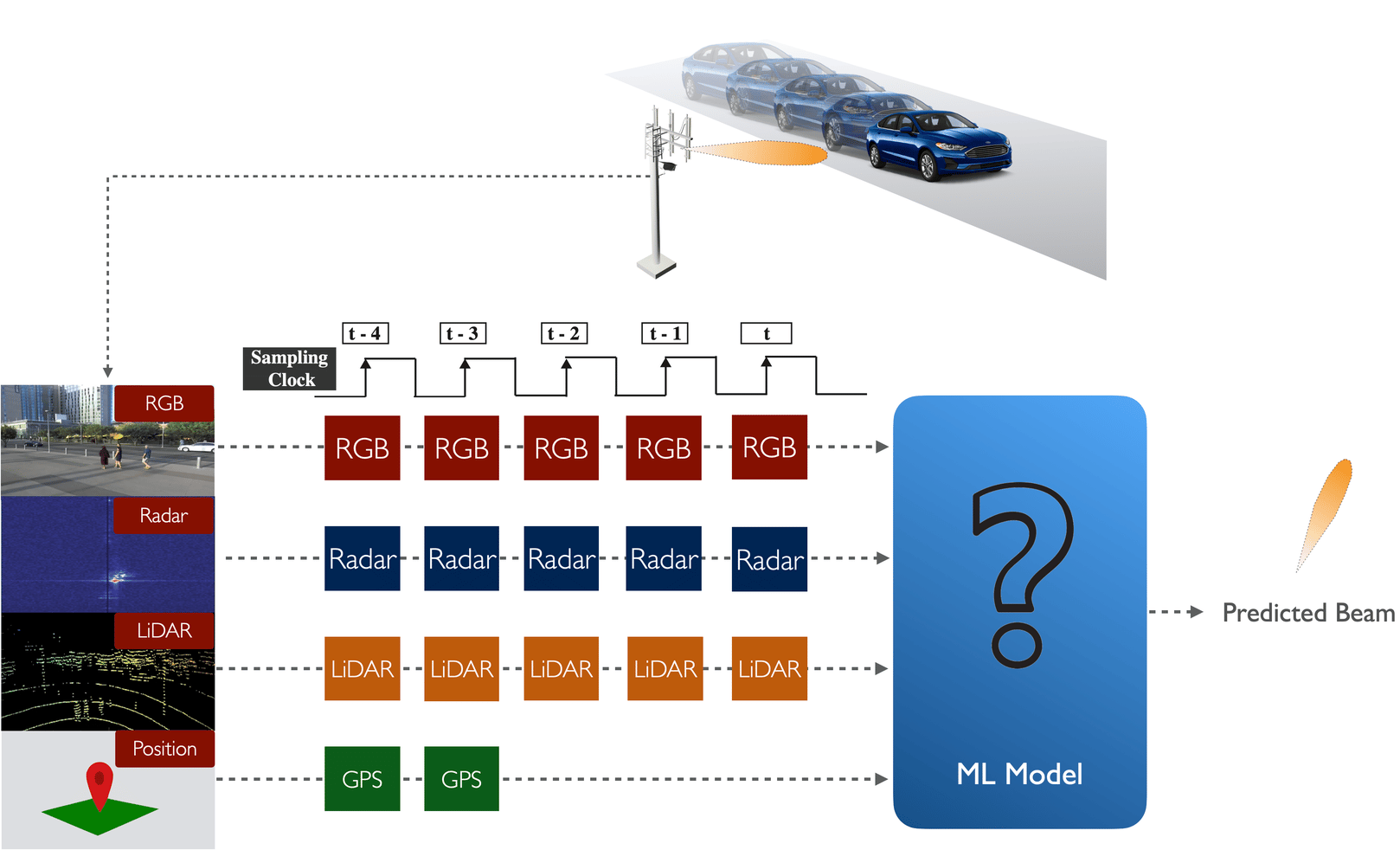}
\caption{Schematic representation of the input data sequence utilized in this challenge task \cite{DeepSense_Challenge}}\label{schematic} 
\end{figure*}

In this paper, we present a transformer-based multimodal deep learning approach for sensing-assisted beam prediction, which is a solution to the DeepSense 6G problem statement in the ITU AI/ML for 5G challenge 2022 \cite{DeepSense_Challenge}. \textcolor{black}{The challenge aims to develop machine learning-based models that can adapt to diverse environmental features and accurately predict optimal beam indices in entirely new locations using a multimodal training dataset. The objective is to enable effective generalization and sensing-aided beam prediction for improved wireless communication systems.} The challenge provides large multimodal sensing datasets measured in a real environment. As shown in Fig. \ref{schematic}, each data sample contains five sequential instances of camera images, LiDAR point clouds, and radar signals, plus the first two instances of UE GPS position \cite{DeepSense}. \textcolor{black}{The ground truth in this context refers to the corresponding 64×1 power vectors obtained through beam training at the receiver using a 64-beam codebook, where omni-transmission is employed at the transmitter. The BS sensors capture LiDAR, radar, and visual data, while the positional data is collected from GPS receivers installed on the mobile vehicle.} The dataset is measured in four scenarios (31, 32, 33, 34) \textcolor{black}{shown in Fig. \ref{mask_pre}. Scenarios 31 and 32 are measured in the daytime while scenarios 33 and 34 are at night. Note that scenarios 32 and 33 are in the same location but at different times.} A development dataset is provided with thousands of samples collected in scenarios 32, 33, and 34; and an adaptation dataset is provided with tens of samples collected in scenarios 31, 32, and 33. Both datasets have the ground-truth best beam of the UE associated with each sample. A test dataset with hundreds of samples is provided in all scenarios without labels. Specifically, most labeled data resides in scenarios 32, 33, and 34, whilst half of the unlabeled data resides in scenario 31. The sampling rate of the sequences in the test set is the same as that of the adaptation set but double that of the development set. The objective is to evaluate how the developed model can generalize to the unseen scenario, in which the sensing data is collected in a different location, Field of View (FoV), time (day, night), and sampling rate.

\textcolor{black}{The evaluation metric is a “Distance-Based Accuracy Score (DBA Score)” with the top-3 predicted beams \cite{DeepSense_Challenge}. The DBA score is defined as 
\begin{align}\label{dba_score}
    \text{DBA score}=\frac{1}{3}(Y_1+Y_2+Y_3),
\end{align}
where $Y_K,\ K\in\{1,2,3\}$ is 
\begin{align}
    Y_k=1-\frac{1}{N}\sum\limits_{n=1}^{N}\min_{1\leq k\leq K}\min\left(\frac{|\hat{y}_{n,k}-y_n|}{\Delta},1\right).
\end{align}
with $y_n$ and $\hat{y}_{n,k}$  are respectively ground truth and the $k$th most-likely predicted beam indices of sample $n$ in the dataset with a size of $N$. $\Delta$ is a normalization factor and set as 5.}


\subsection{Related work}
There exists several solutions for multimodal sensor data fusion for multiple downstream tasks, such as the TransFuser framework proposed for autonomous driving \cite{Prakash2021CVPR}. However, the work is developed for computer vision applications such as semantic segmentation, object detection, recognition, and localization. The data is collected from sensors equipped on the moving vehicles. In comparison, our task has several unique challenges where the TransFuser model is difficult to solve. Firstly, our sensors equipped on the BS produce much wider FoV than those on vehicles. There are many static and moving objects in the scene, but there are no labels or bounding boxes indicating the UE. Secondly, we have also radar signals and GPS location, and how to utilize these modalities to assist our task is unclear. Thirdly, beam prediction is a unique application in wireless communications that has not been well exploited with multimodal sensors. In particular, the relations between radio transmission scenarios and visionary data on abstraction space is not straightforward, making deep learning hard to generalize on unseen scenarios \cite{Nishio2021}. 

The use of visual data for wireless communications has been actively studied in recent years, including most work on beam prediction from the DeepSense group. This includes the use of images for beam tracking with Gated Recurrent Unit (GRU)-based deep learning \cite{Jiang2021}. Radar-aided beam prediction is studied in \cite{Demirhan_DeepSense} using 2D Convolutional Neural Network (CNN). It also proposes FFT to transform radar signals to range angle and velocity maps for CNN. LiDAR-aided beam prediction is investigated in \cite{Jiang_DeepSense}, which is also based on GRU plus an embedding block to convert 3D point clouds to 1D vectors. Position-aided beam prediction is studied in \cite{Morais_DeepSense}, which utilizes Multilayer Perceptron (MLP). A fusion of vision and position has been studied in \cite{Charan_DeepSense}, which concatenates the normalized position with extracted features from CNN to predict beams with MLP. Despite that, a number of solutions have been developed in this domain, most of which are not scalable to different modalities of sensory data. To achieve this we need to build a generalized ML model which can learn the abstracted features between multiple modalities, which is a key target of this paper. 

\subsection{Contributions}
Our contribution can be summarized as follows. First, we develop a multimodal transformer framework for wireless communication application of beam prediction and prove that the model is flexible to adapt to different data modalities in the wireless domain. Second, we investigate several data processing and augmentation techniques in computer vision for wireless applications, alongside model training and validation methods for data imbalance and domain adaptation problems. Third, we validate with real measurement data that our framework is effective in producing beam prediction from multimodal sensory data, and generalize to unseen scenarios. Finally, we discuss that our framework could be extended to a generative model pretrained on sequences of multimodality data and fine-tuned for multiple tasks in radio air interfaces.\footnote{\url{https://github.com/ITU-AI-ML-in-5G-Challenge/DeepSense6G_TII.git}}

The rest of this paper is structured as follows. 
\textcolor{black}{Section 2 describes our developed methods for multimodal sensor data transformation and processing. Section 3 describes our proposed multimodal transformer framework for sensing-assisted beam prediction, with discussions on the training method and extension capabilities.}
Section 4 presents experiments of our solution on the multimodal beam prediction applications with discussions on the studied approaches. Finally, the work is concluded in Section 5 with some future research directions for the framework. 
\section{Multimodal Data Transformation and Processing}

\subsection{Multimodal data transformation}
We start by transforming LiDAR point clouds and radar signals into 2D vector space as well as calibrating GPS location data. 

For LiDAR data, we convert the raw point clouds into an image-like representation through a Bird's-Eye View(BEV). Specifically, the height, intensity, and density of the 3D point cloud are mapped to the red, green, and blue channels of a color image to generate the \textsl{BEV} image. Firstly, the point clouds within the Region Of Interest \textsl{(ROI)} are discretized into grid cells. Secondly, the height and intensity are encoded by their maximum values of the points in each grid cell. Finally, the density of the points is calculated \cite{point_cloud_aug}. The BEV representation for LiDAR point clouds has certain advantages. It can be used on CNN \cite{bev_net} to extract hidden features, which can be further processed with images. Moreover, it can preserve the basic structure of the point clouds and the depth information, while reducing the computational complexity in PointNet \cite{point_cloud_aug}.

\begin{figure}
    \centering
    \setlength{\abovecaptionskip}{0pt}
	\includegraphics[width=\linewidth]{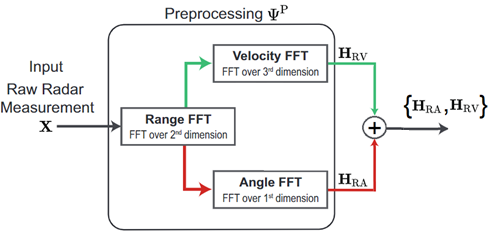}
    \caption{Combining radar Range-Angle $\textbf{H}_{RA}$ and Range-Velocity $\textbf{H}_{RV}$ maps \cite{Demirhan_DeepSense}.}
    \label{radar_pre}
\end{figure}

For radar data, we adopted the processing techniques used in \cite{Demirhan_DeepSense}. The objective is to extract the range, the angles, and the velocity of the moving objects in the environment using \textsl{2D} Fourier transform, as described in \cite{Demirhan_DeepSense}. Since the camera and LiDAR do not provide explicit velocity information, we concatenate the \textsl{Range-Angle Maps} with the \textsl{Range-Velocity Maps} of the radar to preserve the speed information of the moving cars, as illustrated in Fig. \ref{radar_pre}. Further, radar signals provide reliable speed measurement regardless of weather conditions and lightness level \cite{radar_preproc}.

GPS data plays an important role in locating the UE's position. However, it is not always available or accurate in a practical system (caused by connection and delay issues). In this challenge, only data from the first two out of the five GPS instances are provided. We first transform the GPS coordinate of the UE and the BS from to the Cartesian, then calculate the relative position between the UE and the BS of the $n$th GPS data, denoted as $(\Delta x_n,\Delta y_n)$. Afterward, we get the angle by $\arctan(\Delta y_n/\Delta x_n)$.

After exploring the dataset, we observed that the beam indices spread from 1 to 64 according to the UE's locations from left to right in the images. As the camera is located close to the BS, the beam indices are associated with the relative position (angle) between the UE and BS. However, the angles of the same beam index are different between scenarios, because roads are located in different positions with reference to the BS. Therefore, we calibrate the angle of the central pixel in the images of all scenarios.

We first manually select the data samples of these four scenarios where the UE is located in the middle of the images and their corresponding beam indices fall in the range of $[31, 34]$. We then
calculate their angles according to their relative positions, as $[\theta_1 = -50.52^\circ, \theta_2 = 44.8^\circ, \theta_3=55.6^\circ, \theta_4=-60^\circ]$. We rotate all the possible angles in each scenario with $\theta_i$, $(i=1,2,3,4)$. Finally, we obtain the calibrated angles of the first two instances.
\subsection{Multimodal data processing}
\label{sec:data_analysis}
In this section, we introduce several data processing techniques on the multimodal data for training the multimodal transformers on the beam prediction task. 

\subsubsection{Camera data}
Beam prediction from camera data is related to object detection and tracking tasks in computer vision. However, \textcolor{black}{since there are no labels of the targeted UE in the image}, we cannot distinguish it from other vehicles or pedestrians. Therefore, we tried to enhance the visual information of the vehicles in the images to allow the model to better recognize our targeted object.

\begin{figure}
     \centering
     \begin{subfigure}[b]{0.4\textwidth}
         \centering
         \includegraphics[width=\textwidth]{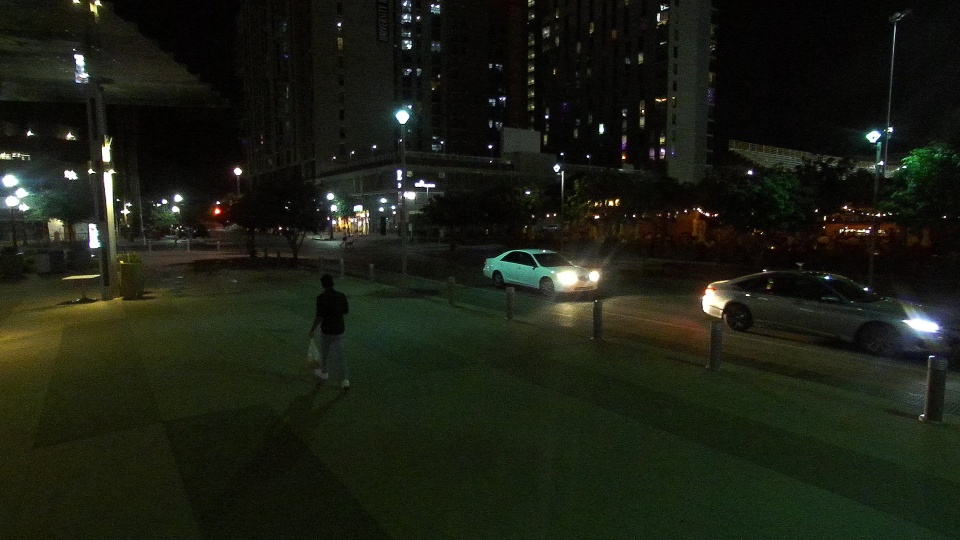}
         \caption{The original image}
         \label{raw}
     \end{subfigure}
     \hfill
     \begin{subfigure}[b]{0.4\textwidth}
         \centering
         \includegraphics[width=\textwidth]{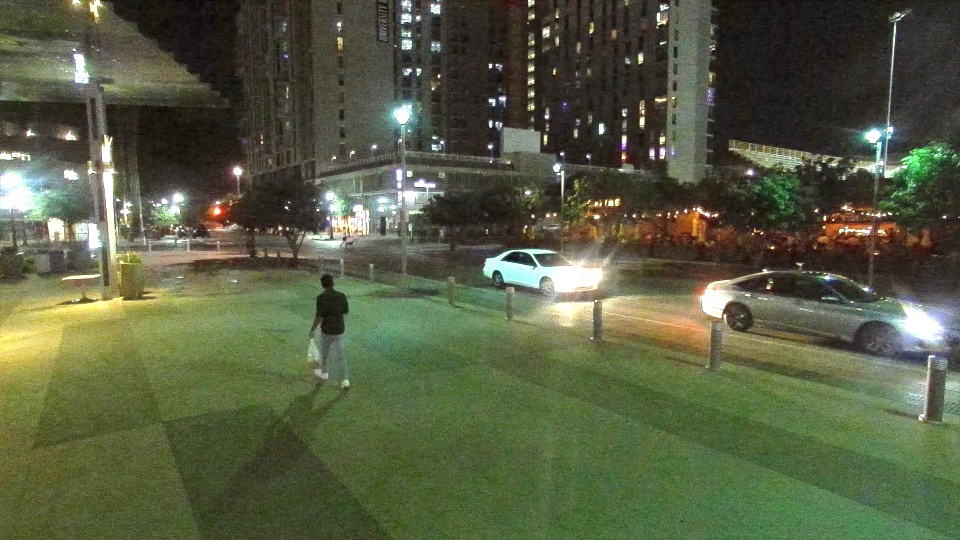}
         \caption{The enhanced image}
         \label{enhanced}
     \end{subfigure}
     \caption{Image enhancement in night scenario}
    \label{enhance_pre}
\end{figure}

\textbf{Brightness enhancement}: To overcome the darkness issue in the night scenarios 33 and 34, we utilize MIRNet\cite{zamir2020learning} to enhance the brightness of these images. The vehicles become clearer as shown in Fig. \ref{enhanced}, compared to the raw image in Fig. \ref{raw}. 

\begin{figure}[ht]
    \centering
    \includegraphics[width=0.4\textwidth]{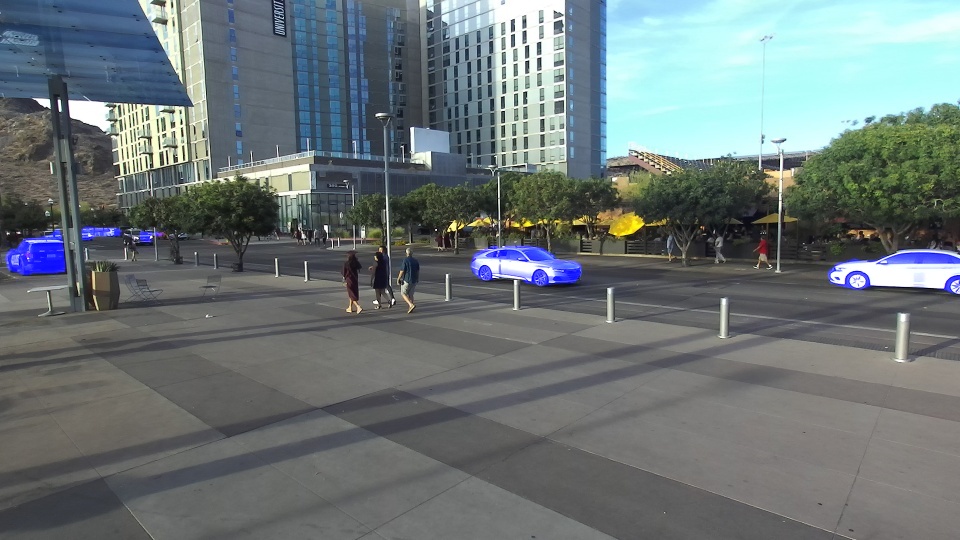}
    \caption{Image segmentation on vehicles (blue) in day scenario}
    \label{segment_pre}
\end{figure}

\textbf{Segmentation}: To highlight the vehicles in the camera data, we use the PIDNet \cite{xu2022pidnet} to segment the vehicles from images in the daytime scenarios 31 and 32 shown in Fig. \ref{segment_pre}. We also test this method on the brightened images in the night scenarios 33 and 34, but the performance is poor, which \textcolor{black}{may be due to loss of background information}. 

\begin{figure}
     \centering
     \begin{subfigure}[b]{0.4\textwidth}
         \centering
         \includegraphics[width=\textwidth]{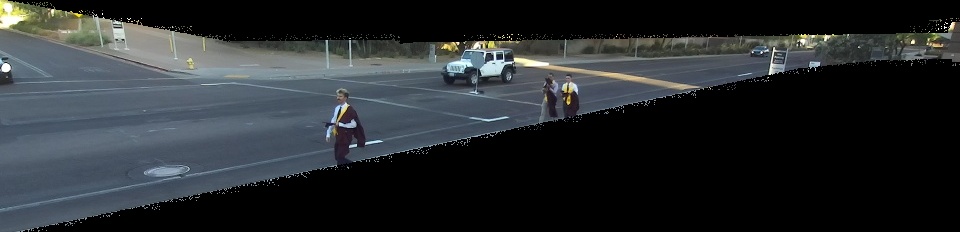}
         \caption{Scenario 31}
         \label{m31}
     \end{subfigure}
     \hfill
     \begin{subfigure}[b]{0.4\textwidth}
         \centering
         \includegraphics[width=\textwidth]{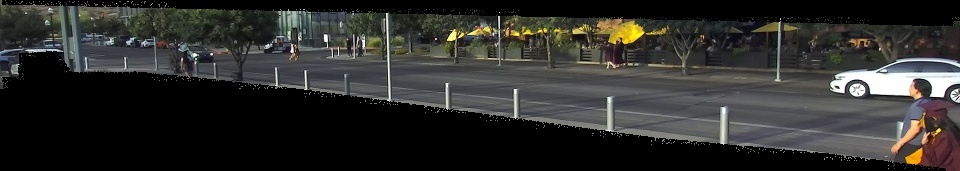}
         \caption{Scenario 32}
         \label{m32}
     \end{subfigure}
    \hfill
     \begin{subfigure}[b]{0.4\textwidth}
         \centering
         \includegraphics[width=\textwidth]{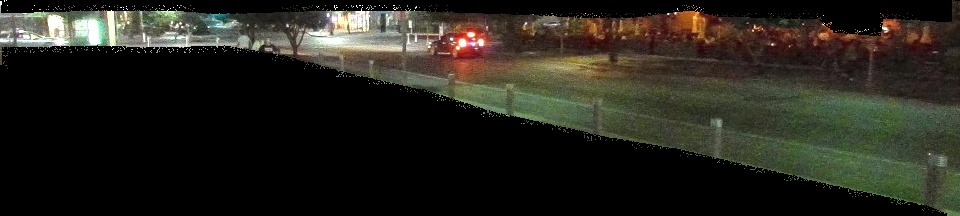}
         \caption{Scenario 33}
         \label{m33}
     \end{subfigure}
    \hfill
     \begin{subfigure}[b]{0.4\textwidth}
         \centering
         \includegraphics[width=\textwidth]{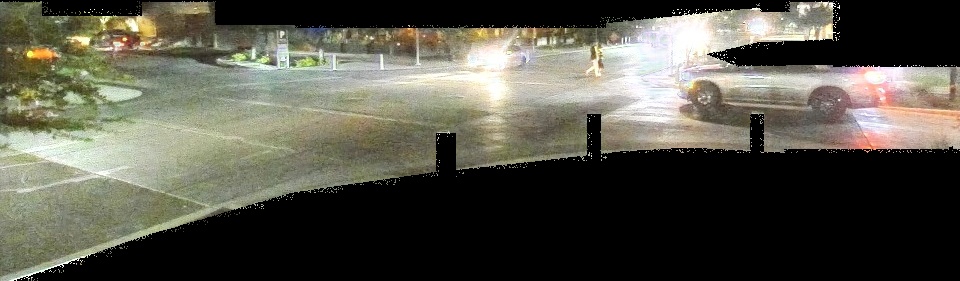}
         \caption{Scenario 34}
         \label{m34}
     \end{subfigure}
    \caption{Image background masking}
    \label{mask_pre}
\end{figure}

\textbf{Background masking}: We also tried to mask the background with the black color and keep the street scene. The images in the same scenario have the same background because the camera is stable. Beam prediction can be partially seen as trajectory prediction over the horizontal axis. We can potentially make the neural network focus on the vehicle's trajectory by making it dominant in the images, as shown in Fig. \ref{mask_pre}.

\subsubsection{LiDAR data} 
The LiDAR produces on average more than \textsl{16000} 3D points in each time step. In order to reduce the size of the point-cloud data to speed up training the model, we preprocessed LiDAR data in the following ways:

\textbf{Background filtering:} We removed data points that correspond to static objects i.e. buildings. Similar to images these regions are not in the Line-Of-Sight \textsl{(LOS)} link between the BS and UE, which has less effect on the beam prediction. These points potentially add complexity and bias to the model. We subtract the background points from each point cloud frame using the moving average of all the frames in each scenario and then keep the desired region surrounding the moving vehicles. 

\textbf{FoV calibration:} We crop the BEV projection of the LiDAR data to keep its FoV consistent with the view in the images. This could potentially assist the CNN focus on the region aligned with the images, to better allow transformers to learn the relations between them. 


\subsubsection{Data augmentation}
Due to data imbalance between scenario \textsl{31} and others, we investigate data augmentation techniques to increase the dataset size for this scenario.

\textbf{Image:} Beam selection relies mainly on the transmitter/receiver locations and the geometry/characteristics of the surrounding environment \cite{Demirhan_DeepSense}. In order to conserve this geometric information, we use only some \textsl{photometric} transformations that are \textsl{'safe'} for beam prediction application \cite{img_aug}. We augment each image by randomly changing the brightness, contrast, gamma correction, hue channel, color saturation, the sharpness, and performing \textsl{Gaussian} blurring on the image.

\textbf{Point-cloud:} Similar to the camera data, we perform two \textsl{'safe'} data augmentation techniques for each point cloud frame without deteriorating the geometric information of the environment: randomly down-sampling the point cloud by a factor of \textsl{10 \%}, and adding small and random 3D position deviation for each point. These transformations conserve the position and general shape of the objects in the environment (cars, buildings, pedestrians, etc).

\textbf{Radar signal:} In order to augment the radar data, we add a small and random noise to each normalized \textsl{FFT} coefficient. The added noise is limited to \textsl{10\%} of each \textsl{FFT} component amplitude in order to conserve the shape of the spectrum. Hence, this transformation is \textsl{'safe'} in the spectral domain. 

\textbf{\textcolor{black}{Multimodal data flipping:}} According to the observations beam indices spread from 1 to 64  in the images, we horizontally flip the images, radar, and the point cloud data to achieve the augmentation purpose. Meanwhile, to keep the GPS data and beam indices consistent with the multimodal data, we reverse the calibrated GPS angles and get the new beam indices by subtracting the original indices from 65.

\section{Multimodal Transformers for Beam Prediction} \label{sec:solution}
In this section, we introduce our solution of a multimodal transformer framework for wireless communications and deep learning algorithms for sensing-assisted beam prediction. 


\subsection{Multimodal transformer architecture}

With multimodality data transformed into 2D vector spaces, we leverage CNN to extract the higher-order features, and then learn the relations between them using transformers. Since the image, point-cloud, and radar signal raw data resides in very different representation spaces, it is difficult to create effective mathematical functions, i.e. through category theory, to transform them into a common abstraction space, in order to learn their structures. However, with multiple layers of learning extraction on CNN and relations on transformers, the deep learning model could potentially converge on an effective representation of multiple modalities. Such representation can be fine-tuned for different downstream tasks which is essentially a structure minimization process. 

\begin{figure*}
	\centering
	\setlength{\abovecaptionskip}{0pt}
	\includegraphics[width=\linewidth]{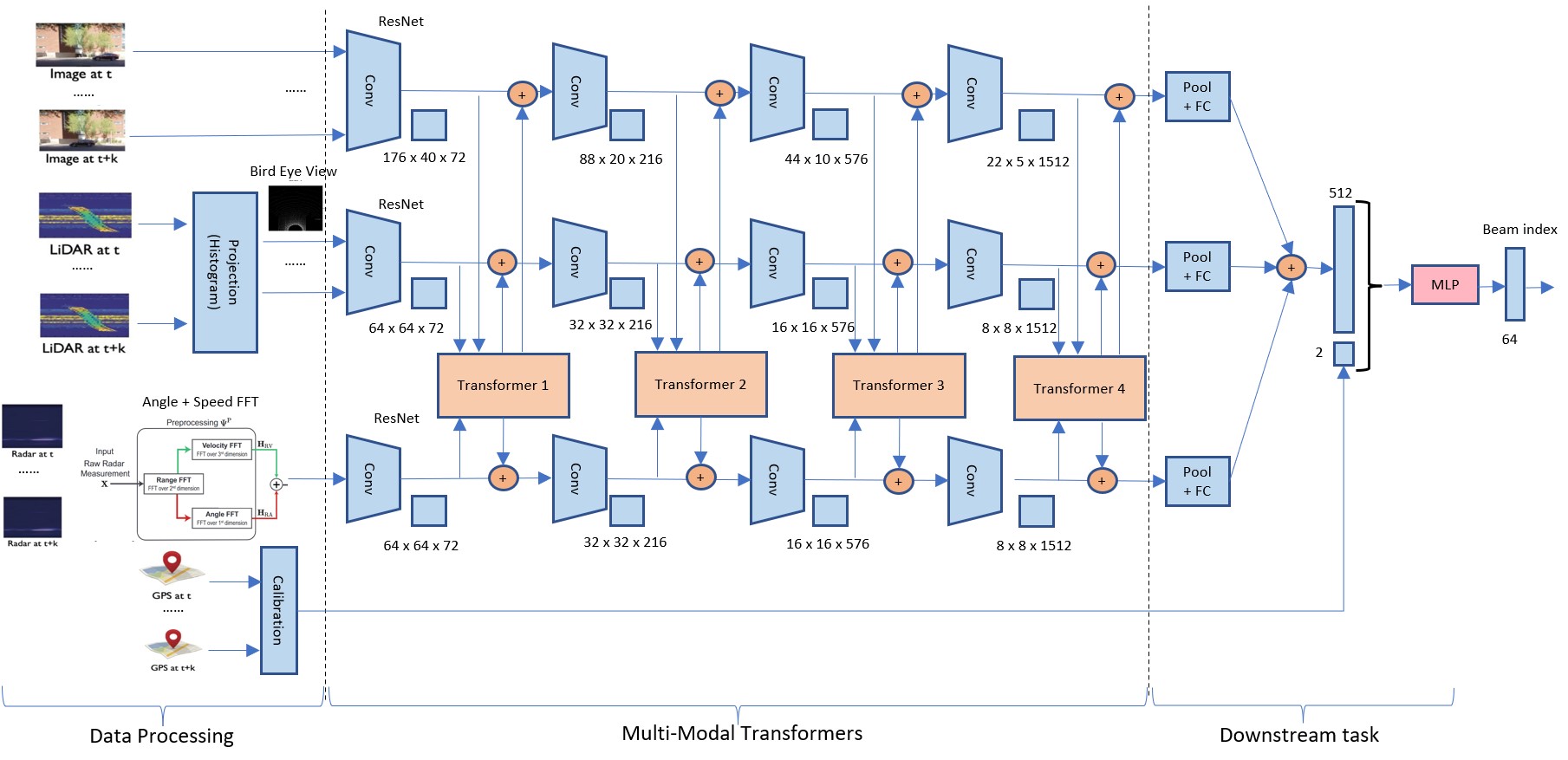}
	\caption{Multimodal Transformers for Sensing assisted Beam Prediction.}
	\label{fig:transfuser}
\end{figure*}

In this context, we build a multimodal transformer architecture as illustrated in Fig. \ref{fig:transfuser}. We first employ a deep residential network (ResNet) \cite{ResNet} to encode the image, point cloud, and radar signal on feature space. Specifically, the ResNet is used on each of the five instances of the RGB image, LiDAR BEV, and radar range angle-velocity map, after normalization and scaling to a $512\times 1$ feature vector. 

Each ResNet block of convolution, batch normalization, non-linear activation, and pooling produces an abstracted feature vector as tokens. Note that for each modality we have five tokens sampled in different time steps. We use transformer encoder layers after each convolutional block to fuse the intermediate abstractions between the modalities of the image, point cloud, and radar map. The transformer uses linear projections for computing a set of queries, keys, and values. Scaled dot products are used between queries and keys to compute the attention weights and then aggregate the values for each query. Finally, a non-linear transformation is used to calculate the output features. It applies the attention mechanism multiple times throughout the structure, resulting in attention layers with multiple heads to generate several queries, keys, and values. Since each convolutional block encodes different aspects of the scene at different layers, thus several transformer blocks are used to fuse these features at multiple scales throughout the encoder. 

The transformer learns the correlation between data at different modalities and time steps. In theory, the fusion of image and point cloud can better represent the scene, especially in some dark and night scenarios. Furthermore, the radar velocity and angle map can position the mobility objects in the scene. 
In this manner, the transformer could estimate the position of the target UE in the scene at the $5^{th}$ instance. 

The fused feature maps of different modalities are propagated to the next convolutional blocks and repeated several times with transformer blocks, and \textcolor{black}{finally added together to be a $512\times 1$ feature vector. Because the calibrated GPS locations (angles) have more apparent information than the other three pieces of data and only the first two instances are available, these two angles are concatenated with the $512\times 1$ vector} and passed through MLP layers to produce weights of 64 beam index using the softmax function. 

\subsection{Training and optimization for beam prediction}

We develop a number of training and optimization mechanisms to customize the model to the beam prediction task. Firstly, we transform the one-hot beam indexes to Gaussian distribution, by positioning the peak at the best beam and cutoff to 0 at its neighboring five beams. This is to adapt the cross-entropy loss function to the DBA score, where higher weights are given if the beams are closer to the best beam. 

We further apply a focal loss \cite{focal_loss} method to improve training on a sparse set of hard examples. Data imbalance is a significant challenge in this task. The data samples from scenario 31 are much less than others. Moreover, some beams have much less probability to be served as the best beam than others. The adaptation dataset is with a different sampling rate than the development dataset. To differentiate between easy and hard examples, a modulating factor $(1 - p_t)^\gamma$ is added to the cross-entropy loss, with tunable focusing parameter $\gamma\geq0$. Intuitively, it reduces the loss contribution from easy examples and extends the range of examples receiving a low loss. 

We also employ several training methods to stabilize the convergence and make the model robust. We maintain the Exponential Moving Average (EMA) of the parameters during training, instead of utilizing the final trained values. This eliminates the fluctuation at the final steps and makes the model robust. 

\section{Performance Evaluations and Discussions} \label{sec:perforamnce}

We performed experiments to train and evaluate our proposed multimodal transformer and data processing frameworks for beam prediction over the DeepSense challenge dataset \cite{DeepSense}. The performance is measured in the DBA score defined in Eq. \eqref{dba_score}, where the distances of the predicted beam to the ground-truth top three beams are averaged according to the mmWave received signal power from the vehicle UE to BS. 

\subsection{\textcolor{black}{Beam prediction accuracy}}

We combine the development and adaptation datasets, then randomly split them into 90\% for training and 10\% for validation (to choose the best model weights and hyperparameters). The learning rate is set to start from $10^{-4}$. We validate and compare the performance using different proposed data preprocessing, augmentation, \textcolor{black}{ResNets (ResNet18 and ResNet34)} and model training approaches, according to the accuracy scores evaluated on the test dataset provided by the organizer. The hyperparameter with the best performance on the validation dataset is submitted for evaluation. Since the training and test datasets have a large imbalance in scenario 31, it can show the generalization capability of the trained model performing in the unseen scenario. 

\begin{table*}[htb]
\scriptsize
    \centering
    \caption{Distance-based Accuracy of Beam Prediction on Multimodal Test Dataset}
    \label{tab:modal}
    \begin{threeparttable}
    \renewcommand{\arraystretch}{0.4}
    \resizebox{\linewidth}{!}{
    \begin{tabular}{ccccccc}
        \hline \\
      \textbf{Data Type\tnote{1}}& \textbf{Scheme\tnote{2}}& \textbf{Overall}& \textbf{Scenario 31}& \textbf{Scenario 32}& \textbf{Scenario 33}& \textbf{Scenario 34} \\ \\
        \hline
     &  \\	
     \multirow{ 8}{*}{Camera}      &\makecell{\textcolor{black}{Raw Image\textsuperscript{18}}} & \textcolor{black}{0.6535} & \textcolor{black}{0.5124}	&\textcolor{black}{0.7457}	&\textcolor{black}{0.7705}	&\textcolor{black}{0.8137} \\
     &  \\ 
     & \makecell{\textbf{Raw Image\textsuperscript{34}}} & \textbf{0.7548} & \textbf{0.6982}	& \textbf{0.7160} & \textbf{0.8024}	& \textbf{0.8494} \\
     &  \\ 
     & \makecell{5th instance\textsuperscript{34}} & 0.6546 & 0.5171	& 0.7568& 0.7548& 0.8204 \\
     &  \\ 
      & \makecell{Brightness Enhancement\textsuperscript{34}} & 0.7327 & 0.6853&0.7469&0.7371&0.8305 \\
     &  \\ 
      & \makecell{Background Masking\textsuperscript{34}} & 0.7571 & 0.6896&0.7383&0.8157&{0.8570} \\
     &  \\ 
      & \makecell{Image Segmentation\textsuperscript{34}} & 0.6979 & 0.5873&0.7556&0.7824&0.8372 \\
     &  \\ 
      & \makecell{EMA Model Weights\textsuperscript{34}} & 0.7146 & 0.6178	&0.7642	&0.7852	&0.8402 \\
     &  \\ 
      & \makecell{Cross Entropy Loss\textsuperscript{34}} & 0.7395 & 0.7018	&0.7420	&0.7410	&0.8234 \\
     &  \\ \hline\\
     \multirow{3}{*}{Radar} 
     & \color{black}\makecell{Range - Angle \& Velocity\textsuperscript{34}} & \textcolor{black}{0.2807} & \textcolor{black}{0.1840}	&\textcolor{black}{0.2827}	&\textcolor{black}{0.4429}	&\textcolor{black}{0.3282} \\
     &  \\
     & \makecell{Range - Angle \& Velocity \textsuperscript{18}} & 0.3563 & 0.2936&0.3160&0.4800&0.3842 \\
     &  \\
     & \makecell{Range - Angle\textsuperscript{18}} & 0.3092 & 0.2462&0.1926&0.4686&0.3313 \\
     &  \\		\hline\\
     \multirow{3}{*}{LiDAR} & \color{black}\makecell{Raw Point-Cloud\textsuperscript{34}} &  \textcolor{black}{0.4362} &\textcolor{black}{0.3171}&\textcolor{black}{0.4037}	&\textcolor{black}{0.6781}	&\textcolor{black}{0.4636}	 \\
     & \\
     & \makecell{Raw Point-Cloud\textsuperscript{18}} & 0.4422 & 0.3260&0.4272	&0.6705	&0.4707 \\
     &  \\
      & \makecell{FoV Calibration\textsuperscript{18}} & 0.4223 & 0.2964	&0.4370	&0.6781	&0.4310 \\
     &  \\
      & \makecell{Background Filtering\textsuperscript{18}} & 0.2794 & 0.2598	&0.2123	&0.2986	&0.3313 \\
     &  \\\hline\\
     \multirow{2}{*}{GPS} & \makecell{\textbf{Angle calibration}} & \textbf{0.7425} & \textbf{0.6353}	& \textbf{0.7704}	& \textbf{0.8229} & \textbf{0.8906} \\
     &  \\
     & \makecell{Angle calibration +\\ distance on 2nd instance} & 0.6266 & 0.4718	&0.6704	&{0.8481}	&0.7262 \\
     &  \\\hline\\
     \multirow{7}{*}{Multimodal} & \makecell{Images\textsuperscript{34} + Radar (Angle)\textsuperscript{18}} & 0.6992 & 0.6304	&0.6938	&0.7533	&0.8010 \\
     &  \\
     &\color{black}\makecell{Images\textsuperscript{34} + Radar (Angle)\textsuperscript{34}} & \textcolor{black}{0.7206} & \textcolor{black}{0.6378}	&\textcolor{black}{0.7383}	&\textcolor{black}{0.8033}	&\textcolor{black}{0.8148} \\
     &  \\
      & \makecell{Images\textsuperscript{34} + Radar (Angle)\textsuperscript{18} \\ + Point-Cloud\textsuperscript{18}} & 0.6356 & 0.5049	&0.7333	&0.7519	&0.7705 \\
     &  \\ 
     & \makecell{\textcolor{black}{Images\textsuperscript{34} + Radar (Angle)\textsuperscript{34}} \\ \textcolor{black}{+ Point-Cloud\textsuperscript{34}}} & \textcolor{black}{0.7358} & \textcolor{black}{0.6649} & \textcolor{black}{0.7938} & \textcolor{black}{0.7919} & \textcolor{black}{0.8142} \\
     &  \\ 
     & \makecell{\textbf{Images\textsuperscript{34} + GPS}} & \textbf{0.7767} & \textbf{0.7253}	&\textbf{0.8000}& \textbf{0.8038} & \textbf{0.8560}\\
     &  \\       
     & \makecell{Images\textsuperscript{34} + GPS \\ (Image Augmentation)} & 0.7127 & 0.5764 & 0.7654 & 0.8576 & 0.8483 \\
     &  \\ 
     & \makecell{\textbf{Images\textsuperscript{34} + GPS} \\ \textbf{(Flipping Augmentation)}} & \textbf{0.7844} & \textbf{0.7298} & \textbf{0.7852} & \textbf{0.8462} & \textbf{0.8433} \\
     &  \\  \hline\\
     \multicolumn{2}{c}{Best score on the leaderboard of the challenge\tnote{3}} & 0.7162 & 0.6536 & 0.7074 & 0.8576 & 0.712\\& \\  
     \hline
    \end{tabular}
    }
    \begin{tablenotes}
        \footnotesize
        \item[1] Data modalities with all 5 (GPS 2) instances unless specified.
        \item[2] Data processing and model training schemes. Focal loss applied in all experiments unless specified. 
        \item[3] Leaderboard: \url{https://deepsense6g.net/ml-task-multimodal-beam-prediction}
        \textcolor{black}{\item[4] The superscript 18 indicates that ResNet18 is used for feature extraction, while 34 denotes the utilization of ResNet34.}
    \end{tablenotes}   
    \end{threeparttable}
\end{table*}

The experimental results of different model training and data processing schemes are shown in Table \ref{tab:modal}. We compare the performances of the model on camera, radar, LiDAR, GPS, and multimodality data. We can first observe that \textcolor{black}{the experiment using ResNet34 to encode images has a higher accuracy than that with ResNet18,} and the experiments with camera data on all five instances of raw images already achieved an overall accuracy of 75\%. This outperforms largely using only the last instance, indicating that the transformer can effectively utilize the relations between images sampled at different times to predict the beams, though car user is not indicated in the image. We can also see that its performance is better than, or similar to, most data preprocessing techniques, such as brightness enhancement, segmentation, and background masking, which further proves that the multimodal transformer model can generalize to different data domains without arbitrary processing. For example, its performance at unseen scenario 31 is close to the same day scenario 32 nearly 70\%, without any data augmentation. Furthermore, we can see that it performs 10\% better in night scenarios 33 and 33, mainly due to the mobility of car lights in the images being easier to identify, than multiple objectives appearing in the day scenarios. 

In the performance of radar and LiDAR data, we can see that the model achieves the lower accuracy than images, and \textcolor{black}{ResNet18 outperforms ResNet34 on encoding these two data}. This is because the radar signals and point clouds received at the BS are reflected by all the moving vehicles and objects, making the model hard to detect the UE. \textcolor{black}{Meanwhile, deeper residual layers lead to overfitting issues.} Moreover, the signal has coverage constraints, causing issues in detecting UEs far away. Specifically, for the radar data, combining range-angle and range-velocity performs 5\% better, which validates that velocity information can help the transformer to predict the UE mobility and select the beam. For the LiDAR data, filtering the background degrades the performance, because reference information of the UE in the environment could be cut out. This also explains that the model with LiDAR performs better than radar which only contains information about moving objects.

In the performance of GPS data, angle calibration on the first two instances achieve the best accuracy in scenario 34 at 89\%, which outperforms the distance and angle calibration on the 2nd instance. This indicates that only two instances of GPS data can predict the beam very effectively, reaching an accuracy of 74\% which is very close to using images. Our angle calibration scheme is very effective while the distance information is less useful. 

The best performance is achieved on multimodal data using images and GPS. It can be observed that the transformer on these two modalities produces an overall accuracy of 77\%, which is better than using them separately. This is much more significant in the unseen scenario 31, with 10\% higher accuracy than using GPS only. This proves the advantage of multimodal fusion on the feature level of our transformer framework. The GPS information can assist the model to identify the UE in images, which improves accuracy in day scenarios. The fusion of images with radar and LiDAR data also largely outperforms using them separately by 30\% to 45\%. \textcolor{black}{Moreover, when it comes to fusing images, employing ResNet34 for encoding radar and LiDAR data yields considerable performance enhancements as compared to using ResNet18}. Furthermore, we also implement the data augmentation techniques in scenario 31. It demonstrates that flipping the images further enhances the performance, reaching the best overall accuracy of 78\% than all other schemes. Finally, we compare our solution with the best score on the leaderboard of the challenge, which uses convolutional autoencoders to fuse the images and GPS data \cite{BeamBench}. It can be seen that our multimodal transformer achieves 7\% better accuracy in overall performance and scenarios 31 and 32, and 13\% better in scenario 34. This further proves the effectiveness of this framework in solving the beam prediction problem. 
\subsection{\textcolor{black}{Model complexity}}
\begin{table}[t]
\scriptsize
    \centering
    \caption{\textcolor{black}{MACs and Params }}
    \label{tab:macs}
    \begin{threeparttable}
    \resizebox{\linewidth}{!}{
    \begin{tabular}{cccc}
        \hline \\
      \textcolor{black}{\textbf{Source}}& \textcolor{black}{\textbf{Block or Method}}& \textcolor{black}{\textbf{MACs}}& \textcolor{black}{\textbf{Params}} \\ \\
        \hline
     &  \\	
     \multirow{ 6}{*}{\textcolor{black}{\makecell{Main blocks\\ in Fig. \ref{fig:transfuser}\\(5 instances)}}}
      & \makecell{\textcolor{black}{ResNet18}} & \textcolor{black}{2,368,733,184}  & \textcolor{black}{11,166,912}\\
     &  \\ 
      & \makecell{\textcolor{black}{ResNet34}} & \textcolor{black}{4,784,652,288}  & \textcolor{black}{21,267,648} \\
     &  \\ 
     & \makecell{\textcolor{black}{Transformer 1} }& \textcolor{black}{127,221,760}  & \textcolor{black}{400,000} \\
     &  \\ 
     & \makecell{\textcolor{black}{Transformer 2}} & \textcolor{black}{506,101,760}  & \textcolor{black}{1,586,432} \\
     &  \\ 
     & \makecell{\textcolor{black}{Transformer 3}} & \textcolor{black}{2,018,836,480}  & \textcolor{black}{6,318,592} \\
     &  \\ 
     & \makecell{\textcolor{black}{Transformer 4}} & \textcolor{black}{8,064,204,800}  & \textcolor{black}{25,220,096} \\
     &  \\ \hline\\
    \multirow{ 3}{*}{\color{black}\makecell{This paper \\in Table \ref{tab:modal}}}&\color{black}\makecell{Overall best scheme   \\ (Images\textsuperscript{34}+GPS)}& \textcolor{black}{34,740,378,624} & \textcolor{black}{54,982,784}	\\
     &  \\
     &\color{black}\makecell{Best scheme of camera  \\data (5th instance\textsuperscript{34})}& \textcolor{black}{6,948,213,248}  & \textcolor{black}{54, 982,272}\\
     &  \\
     &\color{black}\makecell{Best scheme of GPS  \\data (Angle calibration)}& \textcolor{black}{41,472}  & \textcolor{black}{41,920}\\
     &  \\ \hline\\
     \multirow{2}{*}{\textcolor{black}{In \cite{BeamBench}}} & \makecell{\textcolor{black}{Feature extraction}} & \textcolor{black}{191,949,184} & \textcolor{black}{39,998,304}\\
     &  \\	
     & \makecell{\textcolor{black}{Dense model }} & \textcolor{black}{303,616} & \textcolor{black}{304,704} \\
     &  \\\hline\\
    \end{tabular}
    }
    \end{threeparttable}
\end{table}

\textcolor{black}{We investigate the complexity of our proposed framework from aspects of Multiply–Accumulate Operations (MACs) and number of parameters (Params), and then compare ours with the best solution on the leaderboard of the challenge described in \cite{BeamBench}. Note that the best solution of \cite{BeamBench} utilized images, radar, and GPS data. Authors extract features from images by using CNN-based autoencoders. They get a threshold of radar heatmaps using a 2D Constant False Alarm Rate (CFAR), and then apply Density-Based Spatial Clustering of Applications with Noise (DBSCAN) to obtain the object angle. They also calibrate GPS data in a similar way to ours. Finally, these three pieces of preprocessed data are concatenated and go through a dense model to predict the best beam. As we can't get the detailed parameters or the code of CFAR and DBSCAN in \cite{BeamBench}, we only calculate the MACs and Params of the feature extraction and dense model. For our model, we studied the main blocks in Fig. \ref{fig:transfuser} with an input of five-instance data and three of the best schemes in Table \ref{tab:modal}.
From Table \ref{tab:macs}, we observe that the MACs and Params of ResNet34 are nearly double that of ResNet18. MACs and Params of `Transformer 1', `Transformer 2', `Transformer 3', and `Transformer 4' increase quadruply. The most complex block is the 4th transformer. Our overall best scheme with transformers is the one that is most computationally costly. However, when considering only the 5th image, the scheme experiences a substantial reduction of $\frac{4}{5}$ in terms of MACs and Params. On the other hand, the best scheme solely relying on GPS is the simplest among all the solutions. It is important to note that this scheme demonstrates significantly lower complexity while delivering superior performance compared to the best solution presented in \cite{BeamBench}. Therefore, our low-complexity scheme is suitable for scenarios with limited computational resources, making it a viable option. On the other hand, the high complexity scheme is best suited for scenarios that demand high accuracy and possess abundant computational resources.}

\section{Conclusions and Future Work} \label{sec:conclusion}

In this paper, we present a multimodal transformer deep-learning solution for wireless communications and perform a case study in mmWave beam prediction for a target vehicle user. The transformer encoder is used to learn abstracted relations between features of images, point clouds, and radar signals at multiple time instances, extracted by convolutional layers. Multiple layers of transformers and ResNets are stacked to learn higher-level abstractions for downstream tasks. We employed data transformation techniques of point cloud projection, radar range-angle, and range-velocity FFT to convert the multimodal data on 2D vector space, as well as GPS calibration. Furthermore, we develop data processing techniques to improve the task, including image brightness enhancement, segmentation, background masking, LiDAR field of view calibration, and background filtering. We also proposed data augmentation to reduce overfitting in training, including flipping the images. We trained the model by applying focal loss and exponential moving average techniques. 

Experimental results show that our proposed multimodal transformer solution using image and GPS data achieves the best distance-based accuracy of predicted beams at 78.44\%, with effective generalization to each of the scenarios at 73\%, 78.5\%, 84.6\%, 84.3\%, respectively. This outperforms significantly using LiDAR and radar, as well as each single modality. Specifically, the transformer effectively utilizes GPS information to detect the target UE in the images, whilst the images can assist GPS to generalize better in the unseen scenario. Furthermore, it also performs 7\% better than the best state-of-the-art using autoencoders. We can conclude that our proposed multimodal transformer can effectively perform tasks between visual and radio domains, and generalize to different scenarios without customized data preprocessing and augmentation. 

Further advanced deep learning models and techniques are worth studying for improving performance, especially feature extraction from data in other modalities. Domain generalization is an important issue in this task because the data in scenario 31 and the changed sampling rate in the test dataset have a different distribution than the training dataset. The Batchformer \cite{hou2022batch} algorithm is potentially efficient in making the model robust to imbalanced data, by exploring data sample relationships. Moreover, semi-supervised learning such as the FixMatch \cite{sohn2020fixmatch} algorithm can improve the model on unlabeled data by training on pseudo-labels from evaluation confidences. These methods are useful in practice with no additional computing complexity.   

The multimodal transformer framework can be utilized to build a foundation model to empower multiple downstream tasks in wireless communications. We can pre-train with self-supervised learning to build a generative model from sequences of images, LiDAR, radar, and radio signals collected at different times, frequencies, and locations, which learns high-level abstractions and relations among them. The transformer output can be stacked with classification or regression layers and fine-tuned for downstream tasks related to this data, such as channel prediction, beam management, and modulation. The pretrained model can also be used on devices with fewer sensors and less computing power by adapting the model branch and depth. It is a promising research direction to investigate such architecture for foundation models in the wireless communication domain. 

\section*{Acknowledgment}
We would like to thank the support from TII for our participation in the ITU AI/ML for 5G Challenge 2022.  
\printbibliography 

\section*{Authors}
\label{sec:auth}

\begin{wrapfigure}{l}{0.32\columnwidth} 
    \vspace{-.1in}
    \includegraphics[width=0.39\columnwidth]{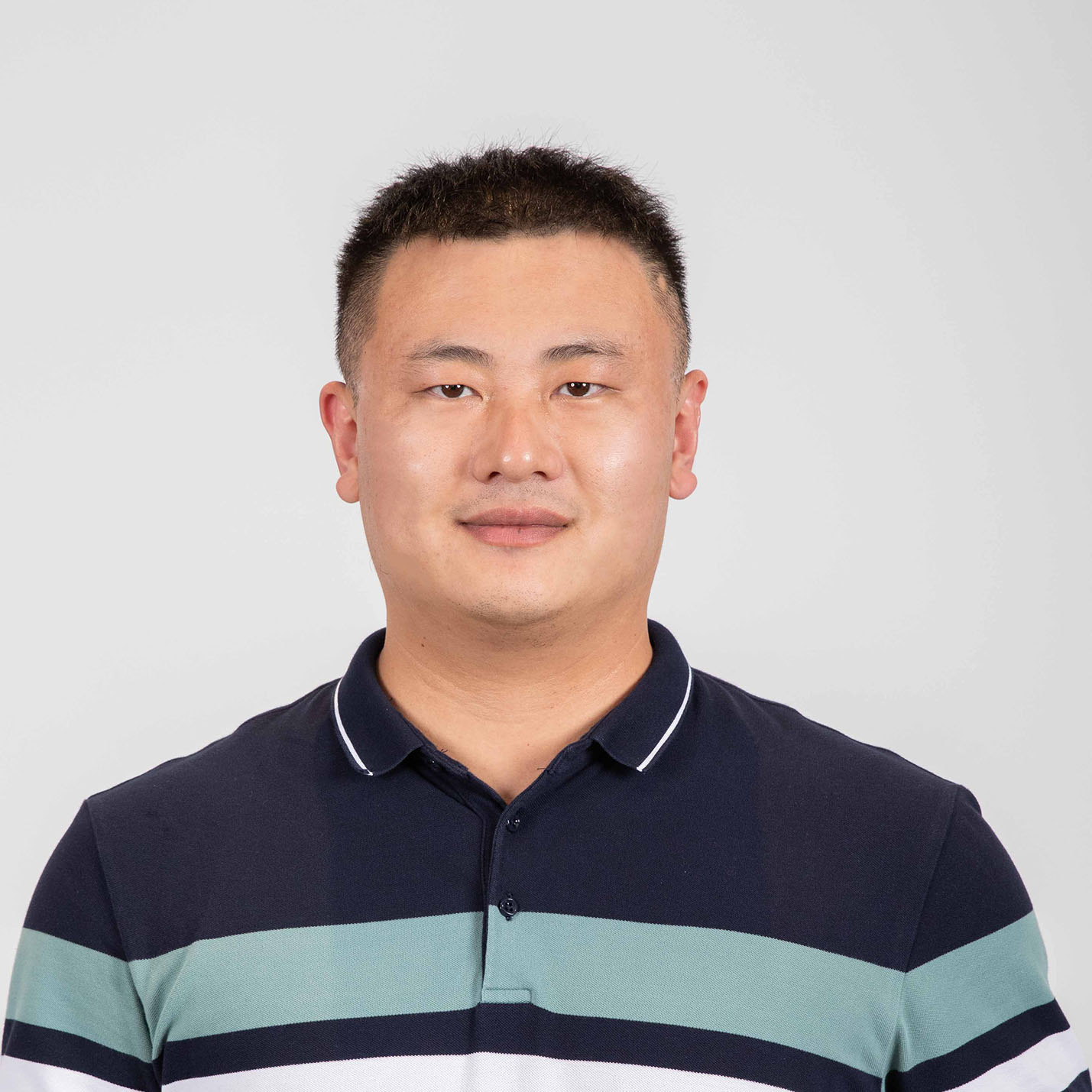} 
\end{wrapfigure}\textbf{Yu Tian}  received a Ph.D. degree from King Abdullah University of Science and Technology (KAUST), Saudi Arabia in 2022. Since June 2022, he has been a researcher at the Technology Innovation Institute, Abu Dhabi, United Arab Emirates. His current research interests include deep learning and performance analysis of wireless communication systems.\ITUpar

\begin{wrapfigure}{l}{0.32\columnwidth} 
    \vspace{-.1in}
    \includegraphics[width=0.39\columnwidth]{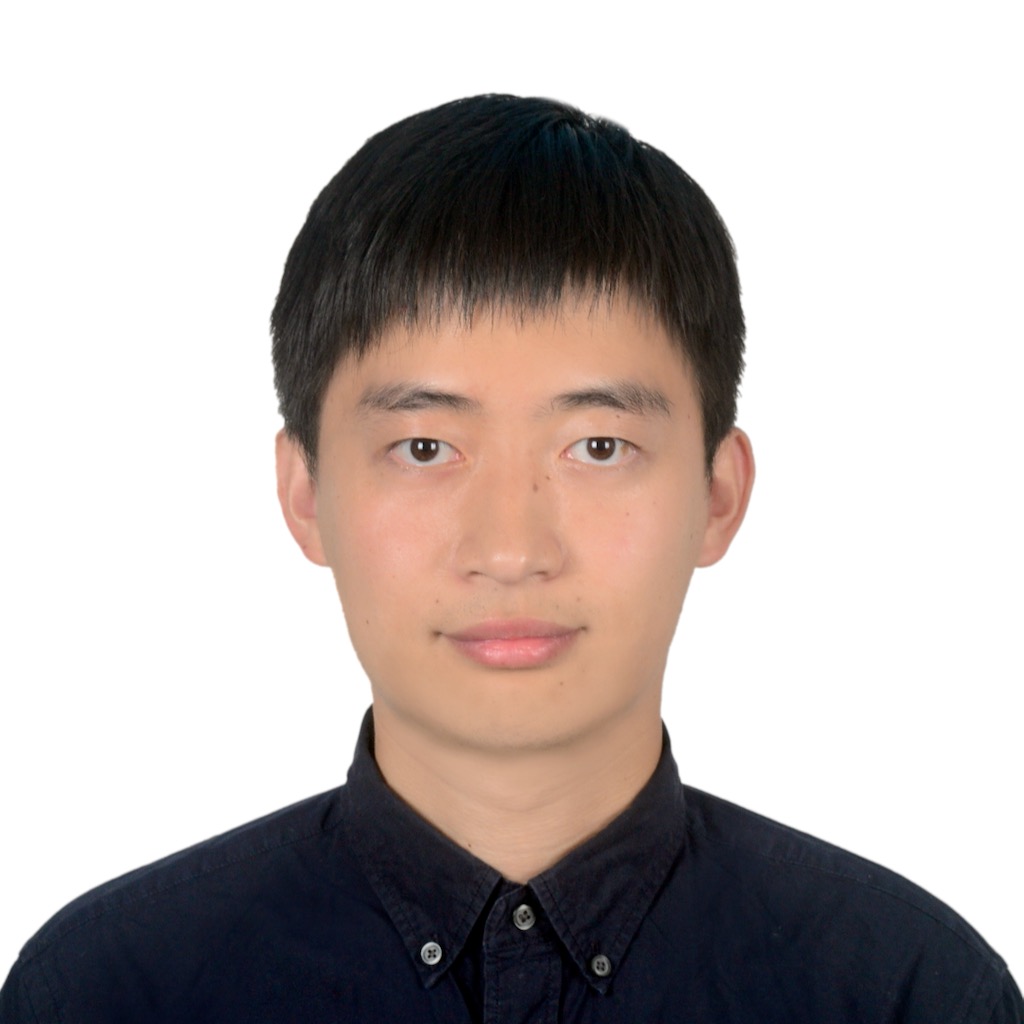} 
\end{wrapfigure}\textbf{Qiyang Zhao} received Ph.D. degree in electronic engineering in 2013 from University of York, UK. He has been working with industry on research, development, and standardization of 5G and 6G wireless communication systems. Since February 2022, he has been with Technology Innovation Institute, UAE, where he is currently a lead researcher. His research interests cover various aspects of AI and telecommunications, with current emphasis on native AI networks, large language models, semantic communications, multi-agent systems. \ITUpar

\begin{wrapfigure}{l}{0.32\columnwidth} 
    \vspace{-.1in}
    \includegraphics[width=0.39\columnwidth]{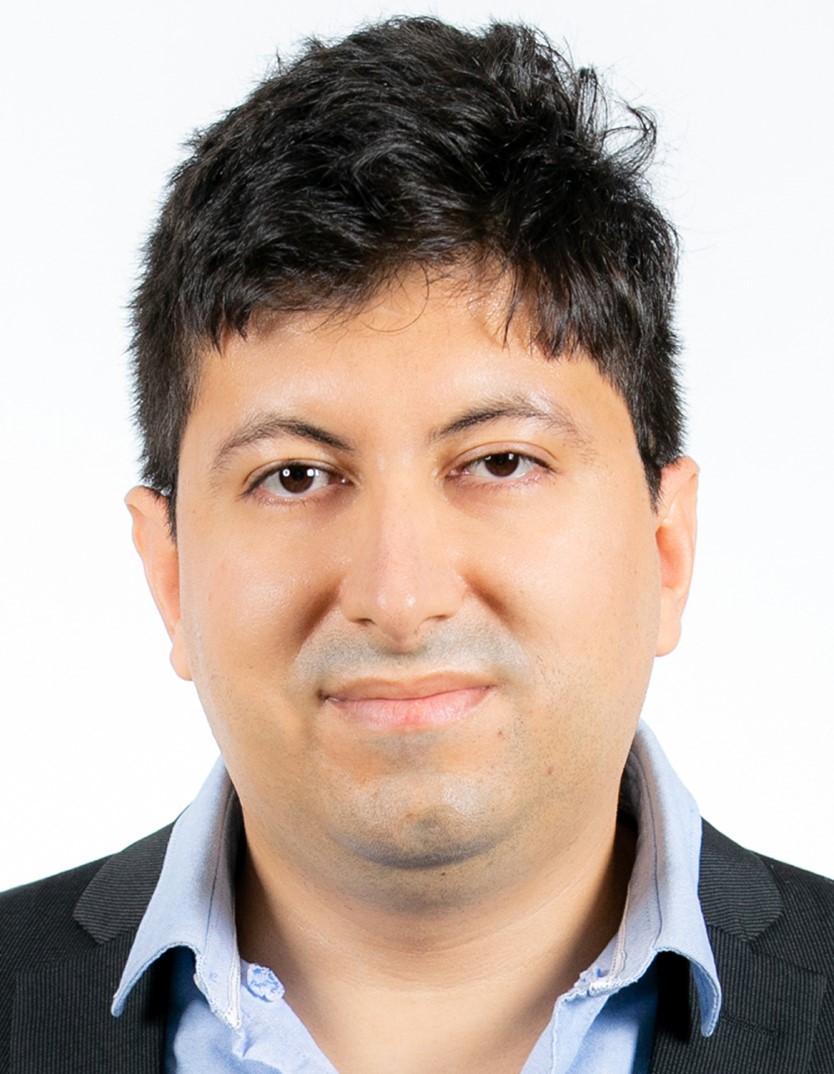} 
\end{wrapfigure}\textbf{Zine el abidine Kherroubi} received a B.S. degree in automatic control systems engineering from École Nationale Polytechnique d’Alger, Algiers, Algeria in 2015 and an M.S. degree in mobility and electric vehicles from Art et Métier ParisTech, Lille, France in 2016, as a Fellow Student of the prestigious Renault Foundation Scholarship. He received a Ph.D. degree in computer sciences from Claude Bernard Lyon 1 University, France in 2020, in joint research between the LIRIS laboratory and Renault car manufacturer. He worked as an R\&D engineer at Renault car manufacturer, France, between 2017 and 2020. He is currently with the Technology Innovation Institute, Abu Dhabi, UAE, as a researcher. His research interests include the application of AI/ML techniques for connected and autonomous vehicles technology, V2X network, collaborative driving, and vehicle-infrastructure cooperation.\ITUpar

\begin{wrapfigure}{l}{0.32\columnwidth} 
    \vspace{-.1in}    \includegraphics[width=0.39\columnwidth]{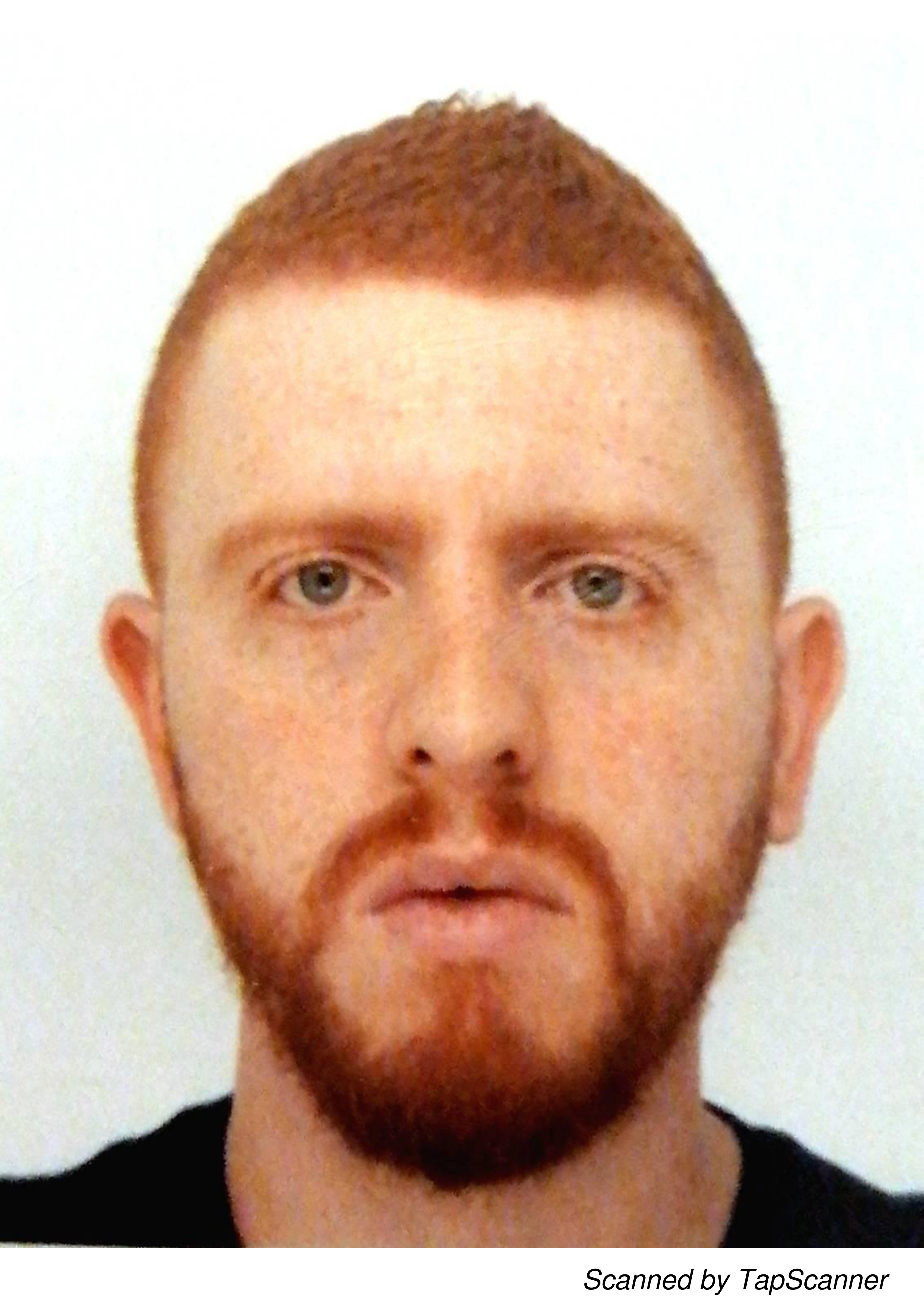} 
\end{wrapfigure}\textbf{Fouzi Boukhalfa} 
received Ph.D. degree from the Sorbonne University, France in 2021, in joint research between INRIA PARIS and VEDECOM. After that, he joined Capgemini Group as a consultant on smart to grid (ENEDIS Nanterre). Since April 2022, he has been a researcher at the Digital Science Research Center in Technology Innovation Institute, Abu Dhabi, United Arab Emirates. His research interests include several aspects of V2X networks for connected and automated driving,  with current emphasis on ISAC and AI solutions for 6G-V2X.\ITUpar

\begin{wrapfigure}{l}{0.32\columnwidth} 
    \vspace{-.1in}
    \includegraphics[width=0.39\columnwidth]{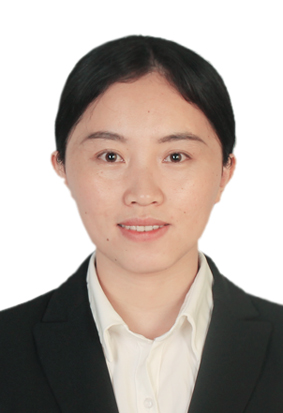} 
\end{wrapfigure}\textbf{Kebin Wu} received a B.S. degree in electronic and information engineering from the Harbin Institute of Technology in 2011 and a Ph.D. degree in information and communication engineering from Tsinghua University in 2018. She is now a senior researcher at the Technology Innovation Institute, with research interests including computer vision, multi-modality learning, and large language models.\ITUpar

\begin{wrapfigure}{l}{0.32\columnwidth} 
    \vspace{-.1in}
    \includegraphics[width=0.39\columnwidth]{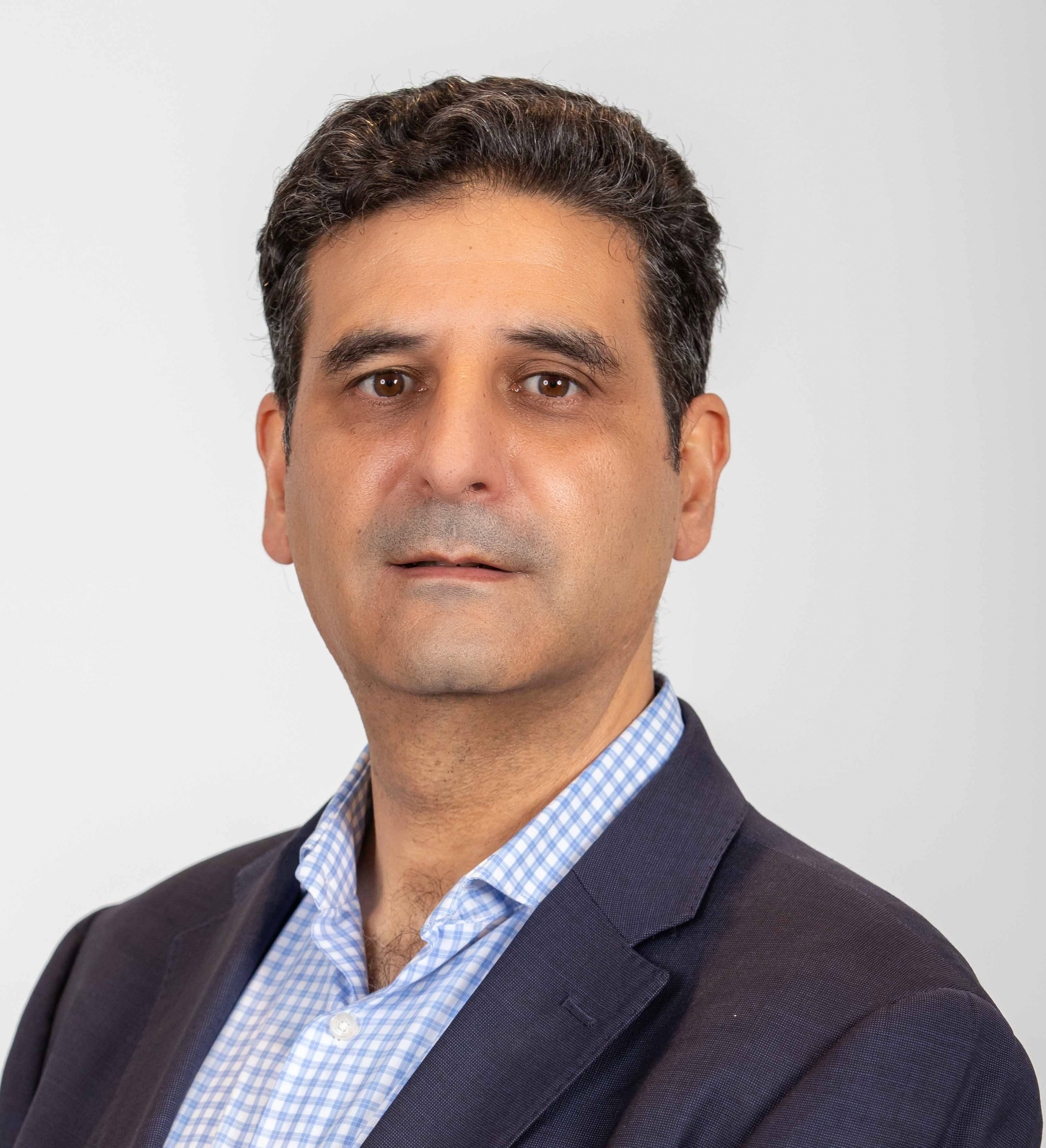} 
\end{wrapfigure}\textbf{Faouzi Bader} received the Ph.D. degree (Hons.) in telecommunications from the Universidad Politécnica de Madrid (UPM), Madrid, Spain, in 2001. He joined the Centre Technologic de Telecomunicacions de Catalunya (CTTC), Barcelona,
Spain, as a research associate in 2002, and
from 2006 to 2013 as a senior research associate. From June to December 2013 he was an associate professor at CentraleSupélec, France. Since 2017 he has been as a honorary adjunct professor with the University of Technology Sydney, Australia, and from 2018 to 2019 as the Head of the Department of Signals and Communications, Institute of Electronics and Digital Technologies (IETR), Rennes, France. From 2020 to 2021 he took up the position of Director of Research at the Institut Supérieur D’Electronique de Paris (ISEP), France. Since December 2021, he has been the Director Telecom at the DSRC Centre of the Technology Innovation Institute (TII), Abu Dhabi, United Arab Emirates. His research interests include IMT-advanced systems such as 5G networks and systems, cognitive radio communication environment, and THz wireless communications (6G). He has been involved in several European projects from the 5th–7th EC research frameworks (eight EU funded projects and ten national projects), he has been the main coordinator of the BRAVE ANR French Project at CentraleSupélec, where the main goal was the achievement of efficient waveform for THz/terabits wireless communication devices. He has published over 45 journals, 136 papers in peer-reviewed international conferences, more than 13 book chapters, and four edited books. He served as a Technical Program Committee Member in major IEEE ComSoc and VTS conferences (ICC, PIMRC, VTC spring/fall, WCNC, ISWCS, GLOBECOM, and ICT).\ITUpar

\end{document}